# A Security Evaluation Framework for Software-Defined Network Architectures in Data Center Environments

Igor Ivkić[1,2], Dominik Thiede[2], Nicholas Race[1], Matthew Broadbent[3] and Antonios Gouglidis[1]
[1]*Lancaster University, Lancaster, UK*
[2]*University of Applied Sciences Burgenland, Eisenstadt, AT*
[3]*Edinburgh Napier University, Edinburgh, UK*
{*i.ivkic,n.race,a.gouglidis*}@*lancaster.ac.uk, 2010781008@fh-burgenland.at, m.broadbent@napier.ac.uk*

Abstract: The importance of cloud computing has grown over the last years, which resulted in a significant increase of Data Center (DC) network requirements. Virtualisation is one of the key drivers of that transformation and enables a massive deployment of computing resources, which exhausts server capacity limits. Furthermore, the increased network endpoints need to be handled dynamically and centrally to facilitate cloud computing functionalities. Traditional DCs barely satisfy those demands because of their inherent limitations based on the network topology. Software-Defined Networks (SDN) promise to meet the increasing network requirements for cloud applications by decoupling control functionalities from data forwarding. Although SDN solutions add more flexibility to DC networks, they also pose new vulnerabilities with a high impact due to the centralised architecture. In this paper we propose an evaluation framework for assessing the security level of SDN architectures in four different stages. Furthermore, we show in an experimental study, how the framework can be used for mapping SDN threats with associated vulnerabilities and necessary mitigations in conjunction with risk and impact classification. The proposed framework helps administrators to evaluate the network security level, to apply countermeasures for identified SDN threats, and to meet the networks security requirements.

## 1 INTRODUCTION

Over the last years, the evolution of cloud computing has triggered a rethinking of the entire Data Center (DC) network architecture (Bilal et al., 2012). One of the main drivers for cloud computing is virtualisation, which enables a massive deployment of computing resources like Virtual Machines (VMs) (Eswaraprasad and Raja, 2017). However, the rise of cloud computing impacts legacy Data Center Networks (DCN) since the number of network endpoints increases, and the Central Processing Unit (CPU) capacity and amount of Random-Access Memory (RAM) of the servers become limiting factors. Additionally, VMs are dynamically created on-demand and can be deleted or moved in the same way. These circumstances bring new requirements to DCN, since (especially in large scale environments) former communication networks and DCs used specialised and costly hardware equipment.

Software-Defined Networks (SDN) aim at providing solutions for these challenges by adding more flexibility compared to traditional DCN. SDN is an emerging architecture that decouples the control plane from the data plane. This separation facilitates the shift of the network control from network devices towards a logically centralised software-based entity called an SDN controller. The SDN controller resides in the control plane, where centralised forwarding tables are calculated and pushed to the network elements living in the data plane. Through the SDN architecture, the network control becomes programmable and provides an abstracted layer where application and network services are applicable. This new solution enables a dynamic, cost-effective, manageable and adaptable network solution matching the increasing network requirements for ambitious cloud applications (Chica et al., 2020). Even though the new SDN architecture opens up new possibilities, it also brings new security challenges.

Compared to traditional network architectures consisting of many distributed forwarding devices the network intelligence is now aggregated in a single virtualised SDN controller. This centralisation of the network control is one of the most significant advantages and, at the same time, the biggest weakness of the SDN topology. A successfully launched cyber-attack where the SDN controller or application is compromised has the potential to affect the entire network (Al-Saghier, 2019). Network security is highly prioritised in DC environments; especially when critical infrastructures are involved, a DC outage resulting from an attack could have disastrous consequences.

In this paper we present a Security Evaluation Framework for SDN architectures in DC environments. Similar to the work of Anisetti et al. (2021), the framework consists of four phases where the output of one phase is used as an input for the next one. In the first phase a threat and vulnerability analysis is performed based on Spoofing, Tampering, Repudiation, Information Disclosure, Denial of Service (DoS), Elevation of Privilege (STRIDE) and Process for Attack Simulation and Threat Analysis (PASTA) (UcedaVelez and Morana, 2015). Next, a risk and impact evaluation is executed for each identified threat and vulnerability, including ranking each risk by its severity by using the Common Vulnerability Scoring System (CVSS). Based on the results of the CVSS, attack scenarios are modelled for the highest-ranked risks. Furthermore, we verify the impact of the modelled attack scenario in an experimental study using a basic SDN setup consisting of Mininet (Kaur et al., 2014) and one Open Network Operating System (ONOS) controller. Finally, we discuss how the identified SDN threats could be mitigated by either using built-in features or by implementing a central security solution.

The remainder of this paper is organised as follows: Section 2 summarises the related work in the field. Next, in Section 3, we present the Security Evaluation Framework and explain how it can be used to evaluate the security of SDN-components. Furthermore, we show the general applicability of the proposed framework in an experimental study in Section 4. Finally, in Section 5 we conclude our work and give an outline of future work in the field.

## 2 RELATED WORK

Since SDN are increasingly becoming more relevant in the cloud computing field, more and more security evaluations and tests have been performed on the new architecture. Ruffy et al. (2016) used the STRIDE model to identify several SDN vulnerabilities and proposed solutions for each of the six security risks to mitigate them. Based on that they sketched a sample of secure SDN design for two network domains. Their design integrated multiple SDN controllers for redundancy together with traditional and new protection solutions, including Intrusion Detection Systems (IDS), firewalls, and access control via Authentication, Authorisation, Accounting (AAA) servers. Nevertheless, they considered just one threat analysis approach based on six generic Threat Categories (TCs), which leads to unrecognised security flaws that are not part of the STRIDE categories.

Iqbal et al. (2019) identified several SDN vulnerabilities and possible attack scenarios including their impacts and focused on possible resolutions for their determined SDN security flaws. Their proposed solution concentrated on the communication channels between the SDN layers instead of considering the entire SDN. Therefore, the resolution proposals are limited to data encryption via Secure Sockets Layer (SSL)/Transport Layer Security (TLS), ciphers like Advanced Encryption Standard (AES) and Data Encryption Standard (DES) and role-based authorisation adapted from FortNOX3 instance. Iqbal et al. (2019) outlined a summary showing the correlation between attack scenario, affected SDN layer/interface, and security. However, in contrast to the approach of Ruffy et al. (2016), the threat analysis and countermeasure proposal from Iqbal et al. (2019) is solely based on a literature review.

Varadharajan et al. (2019) have discussed a policy-based security architecture for SDN and categorised SDN-specific threats into (1) threats against an SDN controller, (2) threats against the networking devices (switches), (3) threats against communication between the controller and the networking devices, and (4) threats against communications between different SDN controllers. Based on those categories they determined five attack scenarios and four security requirements to prevent the successful execution of the attacks. They developed a policy-based security application that resides on the northbound interface of the SDN controller to meet the previously identified security requirements. Furthermore, they analysed the performance of their policy-based security architecture as well as discussed, the capabilities to counteract various security attacks and meet different security requirements using policy-based mechanisms. Even though the approach is promising, their policy-based mechanism provides one option to overcome security flaws in SDN. Considering that one countermeasure does not fix all vulnerabilities, an appropriate solution should include several remediations to increase the security of the entire SDN.

Chica et al. (2020) conducted an SDN security analysis where they highlighted network attacks, threat vectors, and attack surfaces on all three SDN layers and the interfaces in between. In this re-gard, they discussed to what extent SDN can increase network security and how SDN security can be improved. In addition, they briefly discussed potential flaws and inconsistencies still present in the design of security mechanisms for SDN. Moreover, they mentioned potential attack scenarios and their possible impact on the network. However, they did not consider the technical implementation of those attacks and how the network impact can influence their SDN.

Al-Saghier (2019) proposed a framework for evaluating SDN security that is capable of automatically instantiating known attacks against SDN elements. The framework was implemented by adding additional components to the SDN topology, particularly the agent manager, application agent, agent channel, and agent host. Furthermore, misconfiguration, malware, and insider attacks were categorised as attack vectors. Finally, he provided a Proof-of-Concept (PoC) SDN testbed and evaluated the impact on the network when running different attack scenarios.

Sjoholmsierchio et al. (2021) focused on strengthening SDN security by improving the TLS encryption for OpenFlow. According to them, TLS is currently the only mechanism for protecting the control channel and is only considered an optional basis. Based on their claim that TLS is vulnerable to downgrade attacks, they used a protocol dialecting approach to protecting the TLS encrypted OpenFlow protocol against it. In addition to that, they conceptualised a new form of policy-based networking by extending the protocol dialecting functionality into a policy enforcement proxy. Finally, they simulated a downgrade attack in a Mininet-emulated SDN environment and demonstrated how their proposed security mechanism could be used to detect and prevent these attacks. Their results showed a higher security level on the control channel at the cost of increased communication latency by 22% compared to standard TLS.

Shaghaghi et al. (2020) presented a brief overview of the latest security enhancements in the SDN data plane. Furthermore, they identified five characteristics of SDN that can have the most impact on its security. These characteristics include a centralised controller, open programmable interfaces, forwarding device management protocol, third-party network services and virtualised logical networks. They provided an appropriate overview of SDN characteristics prone to exposing security vulnerabilities. In detail, they highlighted the data plane security challenges, solution requirements, and existing solutions.

Jiasi et al. (2019) designed a monolithic security mechanism for SDN based on Blockchain. Their suggested solution decentralises the control plane while maintaining a network-wide view. Furthermore, their mechanism guarantees authenticity, traceability, and accountability of application flows and provides finegrained access control of network-wide resources and a secure controller-switch channel. They achieved this by adding a Blockchain layer between the control and data plane. In addition, they added attribute-based encryption for access control at the northbound interface and a high-performance securely authenticated protocol, called HOMQV, for communication between the Blockchain and the data layer. In conclusion, the Blockchain-based mechanism can be used to record all network flows and events including an implementation of secure protocols with smart contracts.

Kreutz et al. (2013) described several threat vectors that may enable the exploit of SDN vulnerabilities and sketched a design of a secure and dependable SDN controller platform. They identified seven different attack vectors and discussed several mechanisms to address their threats. Their proposed controller platform design introduced three SDN controllers (instead of one) to improve the system's dependability through replication. In addition, their design sketch included the requirement that the switches need to associate with more than one controller dynamically. Moreover, they explained that the replication between the controllers should not rely on a single one to leverage the diversity in the controller platform. Their proposed design to increase the security and dependability of SDN is mainly based on using several SDN controllers which is already applicable in DCNs or other large-scale networks.

Prathima Mabel et al. (2019) gave a brief insight into vulnerabilities of OpenFlow and countermeasures to secure the SDN controller, data plane, and OpenFlow channel. Furthermore, they presented the "flow tracer"-solution for securing the SDN controller from being misused. Their solution can identify and isolate fraudulent flow entries that a hacker may inject. In addition, the flow tracer encrypts the packets towards the data plane using symmetric key encryption. Finally, they simulated a DoS attack in the Mininet emulator to validate their proposed solution. Overall, they showed that this single mechanism can be an effective countermeasure since it can improve the security of the SDN controller, data plane and OpenFlow channel.

Cabaj et al. (2014) highlighted three considerations regarding SDN security and proposed a theoretical solution to detect SDN threats through data mining. The three mentioned SDN drawbacks included (1) limitation of the OpenFlow protocol, (2) centralised network operations, and (3) lack of middleboxes in SDN. Their proposed solution, called Distributed Frequent Sets Analyzer (DFSA), allows

Table 1: Summary of Related Work on Evaluating Security in SDN (template adapted from Scott-Hayward et al., 2013).

| Related Work | Security | | | PoC | SDN Layer/Interface | | | | |
|---|---|---|---|---|---|---|---|---|---|
| | Analysis | Remediation | New Security Solution | | App | App-Ctl | Ctl | Ctl-Data | Data |
| *Ruffy et al. (2016)* | ✗ | ✓ | | | ✗✓ | ✗✓ | ✗✓ | ✗✓ | ✗✓ |
| *Iqbal et al. (2019)* | (✗) | (✓) | ★ | | ✗✓ | ✗✓★ | ✗✓ | ✗✓★ | ✗✓ |
| *Varadharajan et al. (2019)* | | | ★ | ★ | | ★ | ★ | ★ | ★ |
| *Chica et al. (2019)* | (✗) | (✓) | | | ✗✓ | ✗✓ | ✗✓ | ✗✓ | ✗✓ |
| *Al-Saghier (2019)* | (✗) | | | ✗ | ✗ | | ✗ | | ✗ |
| *Sjoholmsierchio et al. (2021)* | | | ★ | ★ | | | | ★ | |
| *Shaghaghi et al. (2020)* | (✗) | (✓) | | | ✗ | ✗ | ✗ | ✗ | ✗✓ |
| *Jiasi et al. (2019)* | | | ★ | ★ | ★ | ★ | ★ | ★ | ★ |
| *Kreutz et al. (2013)* | (✗) | (✓) | ★ | | ✗✓ | ✗✓ | ✗✓★ | ✗✓ | ✗✓ |
| *Prathima Mabel et al. (2019)* | (✗) | (✓) | ★ | ★ | | | ✗✓★ | ✗✓★ | ✗✓★ |
| *Cabaj et al. (2014)* | | | ★ | | | | ★ | | |
| *Fawcett et al. (2018)* | | | ★ | ★ | | | ★ | ★ | ★ |
| **This Paper** | ✗ | (✓) | | ✗ | ✗✓ | ✗✓ | ✗✓ | ✗✓ | ✗✓ |

✗ Security Analysis    ✓ Security Remediation    ★ New Security Solution    () Fully Based on Literature Review

fast detection of attacks that generate massive traffic. DFSA implements different modules in the SDN architecture to facilitate an automatic reaction as soon as a threat is detected which protects the SDN controller against DoS attacks.

Fawcett et al. (2018) introduced a multi-level distributed monitoring and remediation SDN framework for scalable network security called TENNISON. In this regard, they briefly described the main attributes of their proposed security framework, including efficiency and proportionality, scalability and visibility, programmability and extensibility, transparency, availability and resiliency, and interoperability. The TENNISON architecture is based on the ONOS controller and several appliances distributed across the application, coordination, and collection layers in a top-down manner. In addition, TENNISON provides a security pipeline with four flow tables prepended for the other network application tables. This approach achieves network monitoring and remediation without interfering with forwarding functionality and other network services. Fawcett et al. (2018) created a PoC testbed with 350 nodes and 19 switches in Mininet to evaluate the scalability, analysis, and performance of TENNISON using four attack scenarios including DoS, DDoS, scanning, and intrusion.

As shown in Table 1, the identified related work focuses on performing security analyses, remediation or providing new security solutions. Some authors even used a combination of multiple approaches and covered multiple aspects. Various researchers validated their developed solution using a testbed setup, while others argued their suggested solution on a theoretical level. One of the most significant differences of the identified related work is the amount of considered SDN layers/interfaces. For a comprehensive security analysis, all five SDN main components (three layers and two interfaces) should be part of the evaluation. However, several authors concentrated on specific parts of the SDN architecture and did not provide a holistic view including all SDN components. Compared to the identified related work, this paper (1) analyzes threats and vulnerabilities based on the STRIDE and PASTA model on the entire SDN architecture, (2) examines risks and impacts on a DC environment with the aid of CVSS, (3) models attack scenarios for the most critical security flaws and verifies their impact in a PoC testbed, and (4) highlights methods to improve the SDN security. The output of this approach is a comprehensive SDN Security Evaluation Framework, enabling the DC owners to evaluate the security level of their networks including possible subsequent impacts and provide them with remediations to mitigate elaborate security risks. Table 1 provides an overview of the identified related work and draws a comparison to this paper.

## 3 SECURITY EVALUATION FRAMEWORK

The architectural recomposition triggered by SDN brings up exciting opportunities to DCN. Mainly distributed cloud applications hosted in DCs benefit from the emerging solution. However, from a security perspective the SDN topology creates new challenges, since the centralisation of the network control can be a strength and weakness at the same time. More precisely, the entire network is affected as soon as the SDN controller or application is compromised (Al-Saghier, 2019). In this section we present an SDN Security Evaluation Framework and describe its four stages. In this regard, we explain how the framework can be used to identify threats and vulnerabilities, analyse their risks and impact, model possible attack scenarios and apply countermeasures to mitigate them (as shown in Figure 1).

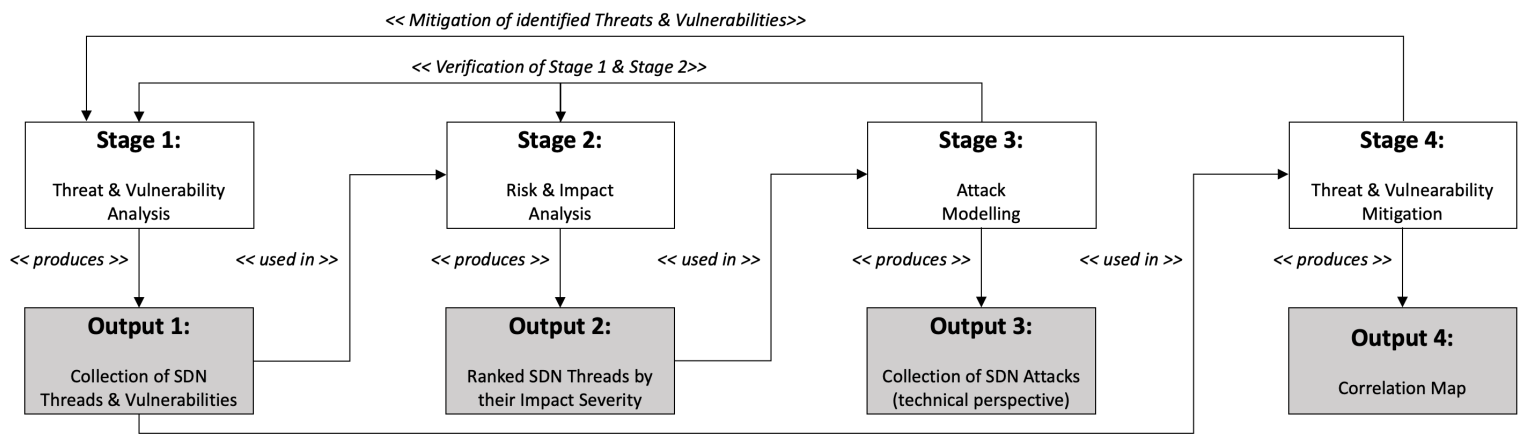

Figure 1: Security Evaluation Framework for SDN Architectures in DC Environments.

## 3.1 Threat & Vulnerability Analysis

Since an SDN controller represents a single point of failure for the entire network, Ruffy et al. (2016) suggest performing a risk-benefit analysis when considering deploying SDN technology. However, before assessing risks, threats must be identified first. Since many threats could be relevant for a security evaluation, threat groups or categories are typically used. The STRIDE model developed by Microsoft is a common approach to distinguish security design flaws on a technical basis and is therefore feasible for the security assessment of SDN (Ruffy et al., 2016).

Categorising threats with the aid of STRIDE is helpful to get an overview of the relevance of certain threat groups. Still, product managers and supervisors are keen to know which threats are relevant to their business, product, and platform. The STRIDE model categories may not include specific threats to SDN. Therefore, a far-reaching risk-based threat modelling methodology, called PASTA also needs to be considered. PASTA integrates business impact, inherent application risk, trust boundaries across application components, correlated threats, and attack patterns that exploit identified weaknesses from the threat modelling. This methodology helps to model SDN-specific threats and assess their business impact. Summarising, the first stage of the Security Evaluation Framework incorporates both the STRIDE and PASTA approaches to evaluate SDN-based security threats and vulnerabilities.

## 3.2 Risk & Impact Analysis

Once the SDN threats and vulnerabilities have been identified in the first step, the next step requires to categorise them based on their impact to the DC environment. Threats with similar impact are then combined to TCs which enables a faster risk and impact analysis. In order to achieve such classification, the outcomes from the STRIDE and PASTA analysis are compared to eliminate possible redundancies. The approach to defining the TCs follows the reverse way compared to the PASTA analysis. During the PASTA assessments, root and sub-threats were determined, and afterwards, attack scenarios and vulnerabilities were derived from the threats. For categorising the security flaws, the effect on the SDN is first defined, and later, the threat correlation occurs.

Next, a risk and impact analysis has to be performed using the Common Vulnerability Scoring System (CVSS). For applying the CVSS scoring system, the National Vulnerability Database (NVD) provides a CVSS calculator which calculates a score in the range between 0 and 10. Based on that score the severity of the TCs can be mapped to one of the following five categories:

Table 2: CVSS Scoring with Severity Base Score Mapping.

| Severity | Base Score Range |
|---|---|
| Critical | 9.0 - 10.0 |
| High | 7.0 - 8.9 |
| Medium | 4.0 - 6.9 |
| Low | 0.1 - 3.9 |
| None | 0.0 |

The CVSS scoring calculates a base score and an overall score for each identified threat. The base score is the foundation for determining the severity, while the overall score represents the severity under consideration of the environmental metric group. In case the overall score is higher than the base score, the TC has a higher impact on the environment as assumed by the threat examination and vice versa. Summarising, the second stage of the Security Evaluation Framework uses a standardised approach (CVSS) to calculate the severity and impact on SDN DC environments of the previously identified threats and vulnerabilities.

## 3.3 Attack Modelling

After identifying threats and vulnerabilities and completing a risk and impact analysis, the third stage focuses on the creation of attack scenarios validating their impact against an SDN testbed. The overall goal at this stage is to validate the results from the risk and impact analysis by modelling attacks for each TC. This could either be done on a real SDN-based DCN, or by creating a close-to-reality test environment (testbed) where the attacks can be executed without disrupting the production environment. Furthermore, in some cases it might make sense to choose between which TCs should be tested based on the calculated CVSS score. For instance, the attacks are modelled for the most likely and most critically ranked threats first before evaluating the other (less likely and less critical) ones. Summarising, the third stage suggests conducting an experimental study where attacks are modelled which provides an additional measure to ensure the correctness of the SDN Security Evaluation Framework.

## 3.4 Threat & Vulnerability Mitigation

The final stage of the framework provides countermeasures that can be applied to mitigate identified threats and therefore improve the overall SDN security. In general, SDN countermeasures can be divided into two main categories that either include methods that mitigate specific threats (e.g., man-in-the-middle attacks) or that increase the security of multiple SDN components and prevent several types of attacks (e.g., Policy-based SDN Security). Mitigations within the first category can be directly mapped to the SDN identified threats from Stage 1 of the Security Evaluation Framework. Such a correlation supports the user of this framework in applying necessary countermeasures to mitigate identified threats and vulnerabilities of their SDN. Those mitigations typically do not require a complex implementation of external tools or software. Instead, the underlying Operating System (OS) or protocols used in SDN environments often provide functionalities to improve security, and it depends on the administrator to enable them.

The second category of mitigations provides a more central solution. Instead of applying countermeasures against each identified SDN threat, a more generic solution can be implemented to secure the network against several attack types of multiple SDN components. These solutions usually go along with a complex and more time-consuming integration of external systems. On the one hand, this approach prevents several threats with one solution. On the other hand, implementing such mitigations requires external systems that have the potential of introducing new vulnerabilities to the SDN. Summarising, the final stage considers both categories to complete the Security Evaluation Framework for SDN in DC environments.

# 4 PROOF-OF-CONCEPT STUDY

## 4.1 Experimental Testbed

For the PoC evaluation a testbed was implemented that simulates the basic functionalities of an SDN in DC environments and follows a simple architecture to reduce any unnecessary functionality overhead. In order to achieve the right balance between simplicity and functionality the testbed uses Mininet for creating a virtual SDN using process-based virtualisation and network namespace features as provided by recent Linux kernels. Mininet simulates links as virtual ethernet pairs which reside in the Linux kernel and connect the switches with the hosts. The Open vSwitch (OVS) is software-based and uses the OpenFlow controller communication protocol which is a key feature for SDN simulation (Mininet, 2022).

Even though Mininet can emulate a rudimentary SDN controller, this controller does not provide enough functionalities for the test network. For this reason, a separate SDN controller is necessary to simulate functionalities from DC environments, for instance, multi-tenancy. The ONOS SDN controller is used as an external controller to manage the switches hosted by Mininet. The combination of the emulated network hosted by Mininet and the ONOS SDN controller provides a suitable base for designing a testbed for the PoC evaluation.

The architecture of the test network is designed as simple as possible but simultaneously complex enough to measure the impact of the different attacks. The test network topology consists of one SDN controller connected to three switches, with links to three VMs each (shown in Figure 2). The SDN applications are not running on a separate machine; instead, the ONOS controller is a platform where SDN applications can be installed directly (ONOS, 2014). One of those applications is Virtual Private LAN Service (VPLS), used in the testbed to isolate multiple domains from each other (Lasserre & Kompella, 2007).

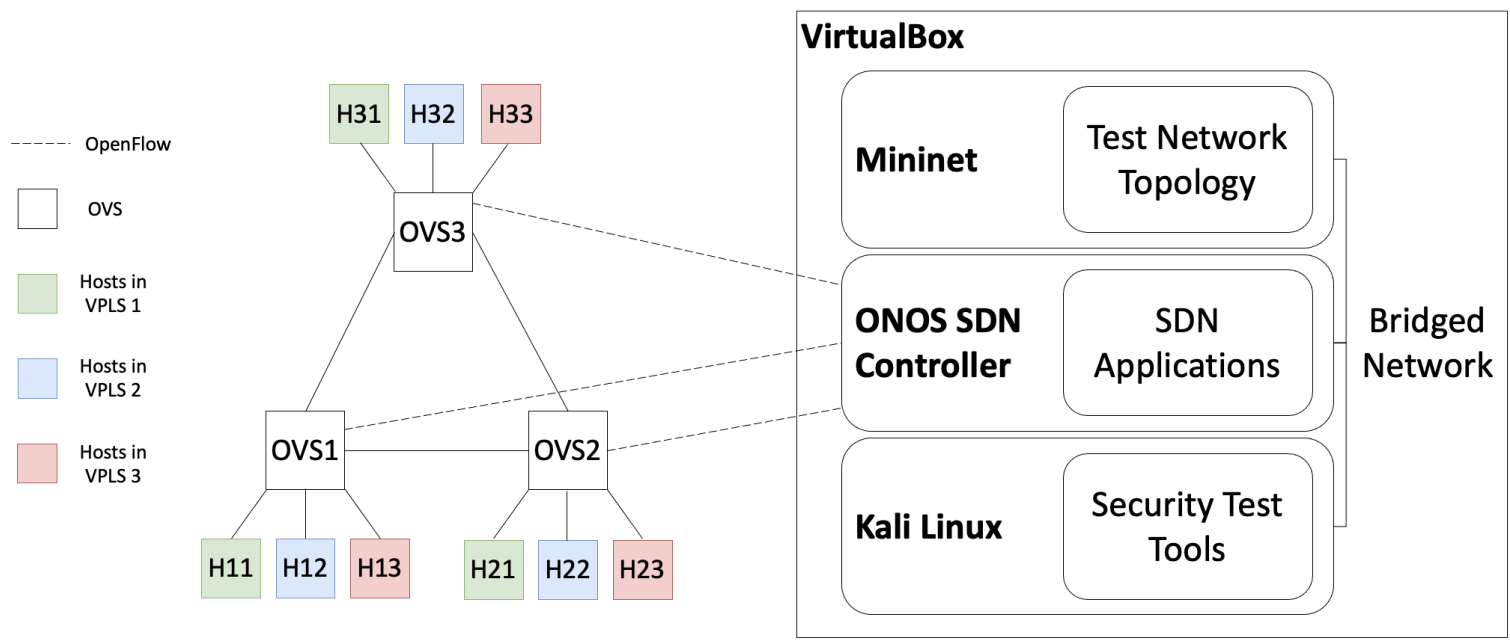

Figure 2: SDN Testbed using Mininet, ONOS Controller and VPLS Services for Network Isolation.

This way, multi-tenancy can be simulated, which is a crucial functionality for SDN in DC environments.

The testbed was configured in a way that hosts can only reach other hosts from the same VPLS. This service is later used in the testbed to verify whether certain attack scenarios can bypass the isolation or reconfigure the service and allow the communication between isolated hosts. In order to run different attack scenarios (Stage 3 of the Security Evaluation Framework) against the testbed, an additional node running Kali Linux is used. Kali Linux is an open-source, Debian-based Linux distribution aimed at advanced penetration testing and security auditing. This provides several tools to test the security in various categories, such as penetration testing, security research, computer forensics and reverse engineering. Due to the comprehensive security tools, Kali is a suitable source to attack the SDN-based testbed.

## 4.2 Experimental Study

As shown in Figure 1, the first stage of the Security Evaluation Framework suggests to perform an SDN Threat & Vulnerability Analysis using STRIDE and PASTA. The main SDN components were investigated separately for all six threats of the STRIDE analysis. Afterwards, the PASTA model was adapted to analyze SDN environments. The modified PASTA model composed of three steps was used to execute an SDN threat and vulnerability analysis. The analysis identified four root threats subdivided into 18 sub-threats and linked to the same amount of vulnerabilities, as summarised in Table 4 and Table 5 in the Appendix (Output 1). Additionally, it turned out that the SDN threats are similar to traditional network threats.

However, due to the SDN architecture's centralisation and the SDN components' dependencies, the same threats have different impacts on the DC environment.

In Stage 2, the identified threats with the same impact on the DC environment were aggregated into categories to eliminate impact redundancies. Next, the CVSS methodology was used to rank the TCs based on their impact severity on a DC environment. As shown in Table 3, the TC with a critical or high severity can potentially harm the entire DC environment and need to be mitigated first. Table 3 shows all 14 TCs, their base and overall score, and severity classes:

Table 3: CVSS Results ordered by TC-Severity (Output 2).

| Rank | | Threat Category | Base Score | Overall Score | Severity |
|---|---|---|---|---|---|
| 1 | TC1 | Unauthorized SDN application access with CSP user permissions | 9,0 | 7,9 | Critical |
| 1 | TC2 | Unauthorized SDN controller access | 9,0 | 7,9 | Critical |
| 2 | TC3 | Man-in-the-middle | 8,9 | 7,9 | High |
| 3 | TC4 | DoS - SDN controller in a single controller setup | 6,8 | 7,7 | Medium |
| 4 | TC5 | Unauthorized SDN application access with tenant user permissions | 6,5 | 5,6 | Medium |
| 4 | TC6 | Unauthorized OpenFlow switch access | 6,5 | 4,6 | Medium |
| 5 | TC7 | Information disclosure of all OpenFlow connections | 5,9 | 6,7 | Medium |
| 5 | TC8 | Information disclosure of the northbound interface | 5,9 | 6,7 | Medium |
| 5 | TC9 | Information disclosure of the BGP connection between controllers | 5,9 | 6,7 | Medium |
| 5 | TC10 | Information disclosure of data traffic | 5,9 | 6,7 | Medium |
| 6 | TC11 | DoS - OpenFlow switch | 4,0 | 2,7 | Medium |
| 6 | TC12 | DoS - SDN application | 4,0 | 3,5 | Medium |
| 7 | TC13 | Information disclosure of a single OpenFlow connection | 3,7 | 2,6 | Low |
| 7 | TC14 | DoS - SDN controller in a multiple controller setup | 3,7 | 2,6 | Low |

In Stage 3, three attack scenarios are modelled for the three highest-ranked TCs from the previous stage. These attack scenarios are then executed on the experimental testbed which has been deployed without any hardening. That means no countermeasures to increase the security of the testbed have been applied. This allows to verify the impact of an attack against the SDN in case security measures are not considered. The following three attack scenarios have been executed (Output 3):

- **Brute Force Attack**: this attack was executed first on the ONOS controller by using four tools on the Kali VM. The fastest tool (Patator) cracked the ONOS default password in 4 seconds, while the slowest (THC Hydra) took $\approx$ 22 minutes.

- **Man-in-the-middle Attack**: this attack was executed on the Mininet VM where packets were captured for the ICMP, Telnet, and OpenFlow protocols. All traffic was analyzed with Wireshark, and it was possible to expose the entire SDN testbed including network services. The same attack also facilitated eavesdropping on an entire Telnet session, including login credentials.

- **DoS Attack**: the DoS attack in the form of an SYN flood attack was executed from the Kali VM against the OpenFlow port (6653) of the ONOS controller. With the aid of the hping3 tool, over 4 Mio. SYN packets were sent to the controller, which caused a traffic interruption between all hosts within their VPLS after 8 seconds and destroyed all three VPLS services. Due to the disrupted network services, the data traffic did not recover, which required a VPLS service reconfiguration to restore the testbed to its initial state.

Finally, in Stage 4 the Security Evaluation Framework provides countermeasures for mitigating the identified and ranked threats and vulnerabilities from the previous stages. For the first category, the mitigations can be directly mapped to the identified SDN threats from Stage 1. In Table 6, we have identified 18 mitigations (M1 – M18) that can directly be mapped to the previously identified threats (T1 – T18) from Stage 1. Moreover, Table 6 provides specific countermeasures per threat based on the MITRE Corporation and OWASP . The presented mitigations (M1 – M18) typically do not require a complex implementation of external tools or software. Instead, the OS or protocols used in SDN environments already provide functionalities to improve security, and it depends on the admin/operator to enable such measures. For instance, the threat (T6) "Network Sniffing" can be mitigated by applying encryption for sensitive data (M6). Data encryption via TLS is a built-in feature of the OpenFlow protocol but is disabled by default. Enabling TLS for the OpenFlow channel increases security. Furthermore, the man-in-the-middle attack from Stage 3 (Attack Modelling) showed how much information about the SDN could be exposed when the OpenFlow protocol is not encrypted.

For threats where the corresponding mitigations are not applicable (M5 and M7), preventive controls are difficult to apply since the threat exploits system features. However, that does not mean that there is no countermeasure present at all. Mitigations in the second category can protect the SDN against such threats by using a central solution. Instead of applying countermeasures against each identified SDN threat, a more generic solution is implemented to secure the network against several attack types on multiple SDN components. These solutions usually go along with a complex integration of external systems.

One of the solutions to mitigate several attack types is the Policy-Based SDN Security Architecture (PbSA) which is a security application implemented in the northbound interface of the SDN controller. The main objective of the PbSA is a security policy-based authorisation of flows in an SDN environment. The security policies use different flow parameters to authorise flows within the SDN. The policy administrator can specify fine-grained security policies based on various attributes such as users, devices, or Autonomous Systems (AS). Furthermore, the PbSA enforces a default deny policy which drops flow requests that are not permitted explicitly. The PbSA contains an additional module to secure the communication between network devices with the aid of distributed keys. A novel feature includes path-based policies establishing an alternative path through the network when the original path is blocked (e.g., DoS) (Varadharajan et al., 2019).

The PbSA mitigates several threats across multiple SDN layers. For instance, it can detect DoS attacks, block suspicious flows and redirect the data traffic over alternative paths. Furthermore, the architecture is capable of blocking all traffic (e.g., dropping malicicious flows or denying misconfigured services) by default that does not satisfy a permit policy which represents an extra security layer in form of an internal SDN firewall. Moreover, the new module ensures a secure communication between several network devices by mitigating the "Network Sniffing"-threat (T6) which enhances the SDN security for the control and data plane, and the southbound interface.

Another central solution was proposed by Weng et al. (2019) which is based on a Blockchain layer between the SDN controller and the data plane. Attribute-based encryption is used to ensure a fine-grained access control for encrypted data at the northbound interface. For the communication between the Blockchain layer and the data layer they establish the HOMQV protocol. The Blockchain layer provides functionalities of resource-sharing and resource-recording among multiple SDN controllers on the control plane. Therefore the Blockchain is used to record the network resources of each controller and to share recorded resources (network events) among all controllers to ensure the same network view.

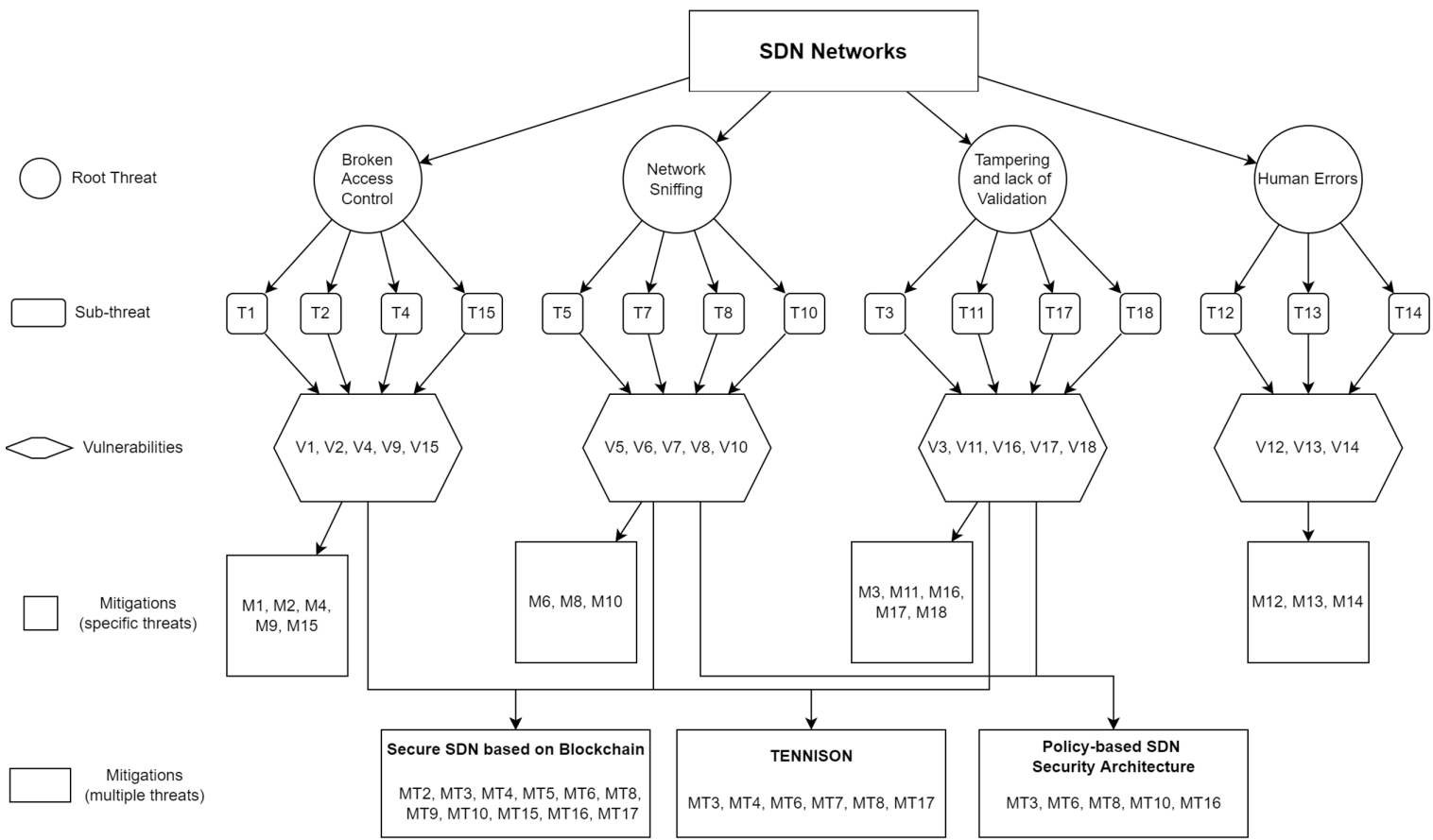

Figure 3: Correlation Map for SDN Threats, Vulnerabilities, and Mitigations (Output 4).

The Blockchain notes all application flows and network events associated with the respective network condition. Such information is stored as a raw transaction in the Blockchain. Furthermore, smart contracts are used to implement security protocols, for instance, to alert the failure of an SDN controller. In addition, the controllers participating in the underlying Blockchain send their recorded network data (from the application and data layer) as raw transactions into the Blockchain. This makes the Blockchain layer a real-time reliability instance of all recorded application flows and all time-series of the network-wide views. The proposed solution provides monolithic security mechanism for each SDN component that can also mitigate threat T5 which cannot be easily mitigated with preventive controls since it is based on the abuse of system features. However, due to the introduced Blockchain layer and HOMQV protocol, this security mechanism provides secure authentication for applications, controllers and switches.

Another way of mitigating several threats with a central solution was developed by Fawcett et al. (2018) where they provided a multi-level distributed monitoring and remediation SDN framework for scalable network security called TENNISON. The TENNISON framework presents an adaptive and extensible security platform that is technology-independent and qualified to support a wide range of security functions. The TENNISON security framework is a comprehensive solution which mainly focuses on the security enhancement of the SDN control and data plane. The distributed ONOS controllers enable availability and resiliency, and the monitoring and remediation functionality detects attacks and drops malicious traffic in the data plane. Fawcett et al. (2018) proved the effectiveness of their solution for four different attack types: DoS, DDoS, scanning, and intrusion. This security framework can also overcome the burdens of mitigating threat T7 which cannot easily be mitigated using specific countermeasures.

Summarising, Figure 3 provides a correlation mapping between the identified threats and vulnerabilities, and the two types of mitigating countermeasures. This mapping enables the user of the framework to find the proper correlation between an SDN vulnerability, the resulting SDN threats and the countermeasures to mitigate them. As shown in Figure 3, there is no single SDN security solution that can mitigate all threats. Therefore, a combination of mitigations against specific threats and a central security solution are required to provide proper security mechanisms for SDN in DC environments.

## 5 CONCLUSIONS

Using SDN in DC environments facilitates the utilisation of cloud applications for multiple users. The emerging architecture enables a dynamic, manageable, and adaptable network solution by decoupling the control from the data plane. However, it also poses new network security challenges.

In this paper we presented an SDN security evaluation framework and showed how it can be used to identify SDN threats and vulnerability including a risk and impact analysis. Once a security threat has been identified, the correlation mapping within the framework directly indicates which vulnerability caused the threat and which countermeasure can be applied to mitigate it. Furthermore, the CVSS scoring shows the impact severity of the threat on the DC environment if countermeasures are not applied. In case a security flaw has been identified which is not mentioned in the evaluation framework, it still provides all necessary models, tools, and procedures to facilitate the extension of the correlation mapping and classification of the impact. The presented framework enables network administrators to evaluate, classify and enhance security of their SDN-based DC networks.

In future work, the SDN threat and vulnerability analysis will be enhanced by adding a complete list of attack scenarios. Due to the enormous number of possible attacks per threat, a complete list would provide more insights into their technical realisation and help to provide mitigating solutions. Furthermore, we plan to increase the accuracy of the CVSS scoring by comparing long and short-term severity impacts on DC environments. We aim at answering the questions, whether attacks that affect the functionality of an SDN for a short period of time are more harmful compared to attacks that longer remain undetected (e.g., eavesdropping on sensitive data). In future work we will provide an answer to this question by introducing an additional metric to the CVSS model to evaluate the impact of long- and short-term effects on DC environments.

# APPENDIX

Table 4: Threat Analysis based on MITRE Corporation and OWASP Top 10 (2021).

| | Threat | Description |
|---|---|---|
| T1 | Command and Scripting Interpreter | • adversaries may abuse command and script interpreters to execute command, scripts, or binaries<br>• attackers may execute commands via terminals/shells or use remote services to achieve remote execution |
| T2 | Modify Authentication Process | • intruders modify the authentication mechanism<br>• access user credentials or enable unwarranted access to accounts<br>• compromised credentials used to bypass access controls placed on various systems and may even be used for persistent access to remote systems and externally available services |
| T3 | Impair Defenses | • maliciously modify components to hinder or disable defensive mechanism<br>• e.g. modify/disable the Access Control List (ACL) for the overlay traffic on the SDN application |
| T4 | Network Boundary Bridging | • attackers may bridge network boundaries by compromising SDN application, controller, or forwarding device<br>• bypass traffic isolation that separates tenant networks<br>• enable movement into new victim environments |
| T5 | Weaken Encryption | • compromising network encryption capabilities to bypass encryption that would otherwise protect data communication<br>• can be achieved by behaviours such as modifying system image, reducing key space, and disabling crypto hardware<br>• greater risk of unauthorized disclosure and data manipulation |
| T6 | Network Sniffing | • sniffing network traffic to capture information about the SDN environment, running services, authentication, etc.<br>• adversaries may set an interface to promiscuous mode to passively access data in transit |
| T7 | Automated Exfiltration | • leverage traffic mirroring to automate data exfiltration over compromised networks<br>• traffic mirroring is a native feature to forward duplicated traffic to one or more destinations for analyzing<br>• adversaries may use traffic duplication in conjunction with network sniffing, input capture, or man-in-the-middle |
| T8 | Active Scanning | • executing reconnaissance scans to gather network information via native features of network protocols such as Internet Control Message Protocol (ICMP)<br>• information from those scans may expose opportunities for other attacks |
| T9 | Broken Access Control | • access control enforces policies to ensure that users cannot act outside of their intended permissions<br>• failures lead to unauthorized information disclosure, modification, or destruction of all data<br>• function execution outside the user its limits |
| T10 | Cryptographic Failures | • identify the protection needs of data in transit<br>• for sensitive data, robust encryption mechanisms need to be in place |
| T11 | Injection | • typical injections are Structured Query Language (SQL), NoSQL or OS command injection<br>• supplied data is not validated by the application<br>• hostile data is used to extract sensitive records |
| T12 | Insecure Design | • broad category representing different weaknesses<br>• missing or ineffective control design<br>• differentiate between design flaws and implementation defects because of different root causes and remediations<br>• a secure design can still have implementation defects, but a perfect implementation cannot fix an insecure design |
| T13 | Security Misconfiguration | • repeatable application security configuration process enables system robustness |
| T14 | Vulnerable and Outdated Components | • keep an overview of the versions of all components<br>• check software version vulnerabilities<br>• regularly vulnerability scans |
| T15 | Identification and Authentication Failures | • validate user its identity, authentication, and session management is critical to protect against authentication-related attacks |
| T16 | Software and Data Integrity Failures | • code and infrastructure that does not protect against integrity violations<br>• application may rely on untrusted sources, repositories, or libraries<br>• attacker could upload their own updates to distribute malicious data across several systems |
| T17 | Security Logging and Monitoring Failures | • without logging and monitoring, breaches remain undetected<br>• systems need to detect, escalate, and respond to active breaches |
| T18 | Server-Side Request Forgery (SSRF) | • occurs whenever a web application is fetching a remote resource without validating the user-supplied Uniform Resource Locator (URL)<br>• enables an adversary to coerce the application to send a crafted request to an unexpected destination |

Table 5: Vulnerability Analysis based on MITRE Corporation and OWASP Top 10 (2021).

| Vulnerabilities | Description |
|---|---|
| V1 | • permission is granted for non-signed scripts<br>• no execution prevention via application control mechanisms<br>• presence of unnecessary or unused shells or interpreters |
| V2 | • lack of multi-factor authentication<br>• lack of privileged account management - ignore the least privilege principle<br>• lack of restricted file and directory permissions |
| V3 | • lack of restricted file and directory permissions |
| V4 | • lack of credential access protection - local passwords may stored in plain-text<br>• lack of user account management - wrong permissions for user accounts<br>• lack of multi-factor authentication<br>• weak password policies |
| V5 | • lack of privileged account management - ignore least privilege principle or using the same credentials across multiple systems |
| V6 | • since the threat is based on the abuse of system features, there is no vulnerability that can easily mapped<br>• lack of multi-factor authentication |
| V7 | • lack of data encryption |
| V8 | • lack of traffic monitoring |
| V9 | • violation of principle of least privilege or deny by default<br>• bypassing access control checks by modifying the URL<br>• accessing API with missing access controls<br>• permitting viewing or editing someone else its account, by providing its unique identifier |
| V10 | • data transmitted in clear-text by using protocols like HTTP, SMTP, File Transfer Protocol (FTP)<br>• old or weak cryptographic algorithms or protocols<br>• encryption not enforced<br>• lack of validation of received server certificate and the trust chain<br>• usage of deprecated hash functions such as Message-Digest Algorithm 5 (MD5) or Secure Hash Algorithm 1 (SHA-1) |
| V11 | • user-supplied data is not validated, filtered, or sanitized by the application<br>• dynamic queries are used directly in the interpreter |
| V12 | • lack of business risk profiling<br>• failure to determine what level of security design is required |
| V13 | • missing appropriate security hardening<br>• unnecessary features are enabled or installed<br>• default accounts and their passwords are still enabled and unchanged<br>• the latest security features are disabled or not configured<br>• the software is out of date or vulnerable |
| V14 | • no overview of the versions of all components<br>• software is vulnerable, unsupported, or out of date<br>• lack of regularly vulnerability scans<br>• lack of upgrading the underlying platform<br>• lack of compatibility tests after updated, upgraded, or patched libraries<br>• security misconfiguration on components (as mentioned in Security Misconfiguration) |
| V15 | • application permits brute force or other automated attacks<br>• usage of default, weak, or well-known passwords, such as "admin/admin"<br>• usage of weak or ineffective credential recovery and forgot-password processes<br>• usage of plain text or weakly hashed passwords data stores (as mentioned in Cryptographic Failures))<br>• missing multi-factor authentication<br>• exposition of session identifier in the URL<br>• reuse session identifier after successful login |
| V16 | • application relies upon plugins, libraries, or modules from untrusted sources or repositories<br>• lack of sufficient integrity verification for auto-update functionality |
| V17 | • lack of logging auditable events, such as logins and failed logins<br>• unclear log messages<br>• logs of applications and APIs are not monitored for suspicious activity<br>• logs are only stored locally<br>• alerting thresholds and response escalation processes are not in place<br>• application cannot detect, escalate, or alert for active attacks in real-time or near real-time |
| V18 | • application is fetching a remote resource without validating the user-supplied URL |

Table 6: Mitigations for identified SDN Threads (T1 – T18).

| Mitigations | Description |
|---|---|
| M1 | • only permit execution of signed scripts<br>• remove any unnecessary or unused shells or interpreters<br>• use application control where appropriate |
| M2 | • enable multi-factor authentication<br>• apply password policies according to Grassi et al. (2017)<br>• ensure proper privilege separation |
| M3 | • restrict file and directory permissions |
| M4 | • ensure proper user permissions are in place<br>• enable multi-factor authentication<br>• apply password policies according to Grassi et al. (2017)<br>• restrict administrator accounts to as few individuals as possible |
| M5 | n/a* |
| M6 | • enable multi-factor authentication |
| M7 | n/a* |
| M8 | • encrypt sensitive information, e.g., with SSL/TLS |
| M9 | • minimize the amount and sensitivity of data available to external parties<br>• implement access control mechanisms like ACLs<br>• log access control failures<br>• rate limit access to minimize the harm from automated attack tooling<br>• identify sensitive data according to privacy laws, regulatory requirements, or business needs<br>• encrypt all data in transit with secure protocols such as TLS<br>• do not use legacy protocols such as File Transfer Protocol (FTP) and Simple Mail Transfer Protocol (SMTP) for transporting sensitive data |
| M10 | • use server-side input validation<br>• use SQL controls like LIMIT to prevent mass disclosure |
| M11 | • use threat modeling for critical authentication and access control<br>• limit resource consumption by user or service |
| M12 | • establish a repeatable hardening process in an automated manner across multiple systems<br>• use ACLs to provide effective and secure separation between components or tenants |
| M13 | • remove unused dependencies, unnecessary features, components, and files<br>• continuously inventory the versions of all components<br>• only obtain components from official sources over secure links |
| M14 | • enable multi-factor authentication<br>• do not use default credentials<br>• apply password policies according to Grassi et al. (2017)<br>• limit or increasingly delay failed login attempts and log all failures<br>• use digital signatures to verify the data is from the expected source<br>• ensure libraries are consuming trusted repositories |
| M15 | • ensure that there is a review process for configuration changes |
| M16 | • ensure all login and access control failures can be logged with sufficient user context<br>• ensure log data is encoded correctly to prevent injections or attacks |
| M17 | • enforce "deny by default" firewall policies or network access control rules<br>• sanitize and validate all client-supplied input data<br>• do not send raw responses to clients |
| M18 | |

*This threat is difficult to mitigate with preventive controls since it is based on the exploit of system features